\documentclass[nalefd]{achemso}
\AtBeginDocument{\nocite{achemso-control}}
\usepackage{achemso}
\usepackage{graphicx}
\usepackage{amsmath}
\usepackage{amssymb}
\usepackage{bm}
\usepackage{dcolumn}
\usepackage{subfigure}
\usepackage{units}
\usepackage{hyperref}
\usepackage{longtable}
\usepackage[usenames]{color}
\usepackage{multirow}
\usepackage[sort&compress,numbers,super]{natbib}
\hypersetup{pdfborder=0 0 0,colorlinks=true,citecolor=blue,linkcolor=blue}
\input epsf

\newcommand{\BN}{\textit{h}-BN}

\newcommand{\MS}{MoS$_2$}
\newcommand{\MX}{MX$_2$}
\newcommand{\MXP}{$\mathrm{M}'\mathrm{X}'_2$}

\newcommand{\MSe}{MoSe$_2$}
\newcommand{\MTe}{MoTe$_2$}
\newcommand{\WS}{WS$_2$}
\newcommand{\WSe}{WSe$_2$}
\newcommand{\WTe}{WTe$_2$}
\newcommand{\NbS}{NbS$_2$}
\newcommand{\TaS}{TaS$_2$}

\newcommand{\MO}{MoO$_3$}
\newcommand{\WO}{WO$_3$}

\title{Ohmic contacts to 2D semiconductors through van der Waals bonding}
\author{\it{Mojtaba Farmanbar},}
\email{m.farmanbar@utwente.nl}
\author{\it{Geert  Brocks}}
\affiliation{Faculty of Science and Technology and MESA+ Institute for Nanotechnology, University of Twente, P.O. Box 217, 7500 AE Enschede, The Netherlands}

\begin{document}
\nocite{achemso-control}
\begin{abstract}
High contact resistances have blocked the progress of devices based on \MX\ (M = Mo,W; X = S,Se,Te) 2D semiconductors. Interface states formed at \MX/metal contacts pin the Fermi level, leading to sizable Schottky barriers for p-type contacts in particular. We show that (i) one can remove the interface states by covering the metal by a 2D layer, which is van der Waals-bonded to the \MX\ layer, and (ii) one can choose the buffer layer such, that it yields a p-type contact with a zero Schottky barrier height. We identify possible buffer layers such as graphene, a monolayer of \BN, or an oxide layer with a high electron affinity, such as \MO. The most elegant solution is a metallic \MXP\ layer with a high work function. A \NbS\ monolayer adsorbed on a metal yields a high work function contact, irrespective of the metal, which gives a barrierless contact to all \MX\ layers.    
\end{abstract}

\maketitle
%************************************************INTRODUCTION*****************************************

\section{Introduction}
Layered transition metal dichalcogenides \MX, M = Mo,W, X = S,Se,Te, are widely explored because of their unique properties and their potential for applications in electronic devices.\cite{Chhowalla:natchem13,Cui:advmat15} \MX\ monolayers are direct band gap semiconductors with band gaps in the range of 1-2 eV, which have appealing electronic and optoelectronic properties.\cite{Wang:nnano12,Zhang:advmat13} \MX\ layers can be obtained via micro-mechanical cleaving,\cite{Geim:nat13} by chemical vapor deposition (CVD),\cite{Lee:advmat12,Chen:advmat15} or even by spin coating precursor molecules.\cite{George:afm14} Important for applications in devices is the ability to have both electron (n-type) and hole (p-type) transport in these 2D materials. Charge carrier transport in \MX\ field-effect transistors (FETs) is usually dominated by electrons; p-type transport has only been demonstrated in \WSe.\cite{Fang:nanol12,Shokouh:afm15} 

A major challenge for p-type transport is that \MX\ forms a large Schottky barrier (SB) for holes with metals commonly used for making electrical contacts. A standard way to reduce a metal/semiconductor contact resistance is to heavily dope the semiconductor in the contact region, which effectively decreases the SB width. Local doping of a 2D semiconductor is however very challenging; so far most techniques used for doping 2D materials, such as substitutional doping,\cite{Joonki:nanol14} adsorbed molecules\cite{Fang:nanol12,Kang:afm15,Cakir:jmcc46}, or electrolytes\cite{Allain:acs8,Braga:nanol12}, have a limited spatial resolution. Alternatively one tries to decrease the SB height, essentially by covering the metal by a thin layer to increase its work function. Oxides have shown their potential for p-type contacts in organic photovoltaics and light-emitting diodes,\cite{Greiner:afm13,Kroeger:apl09} and have also been tested in \MS\ FETs.\cite{Chuang:nanol13,McDonnell:acs8} Oxides have also been applied succesfully to reduce the SB height for n-type contacts to \MS.\cite{Chen:nanol13}

Common metals generally give n-type contacts with substantial SB heights, leading to high contact resistances. Although \MX\ monolayers are free of dangling bonds, nevertheless they interact with low work-function metals to form a density of interface states with energies inside the \MX\ band gap, which is sufficiently large to pin the Fermi level and cause a sizable SB for electrons.\cite{Farmanbar:prb15,Gong:nanol14,Kang:prx14} We show that high work-function metals yield high SBs for holes by a similar mechanism. 

\begin{figure}[b]
\includegraphics[width=9cm, angle=0]{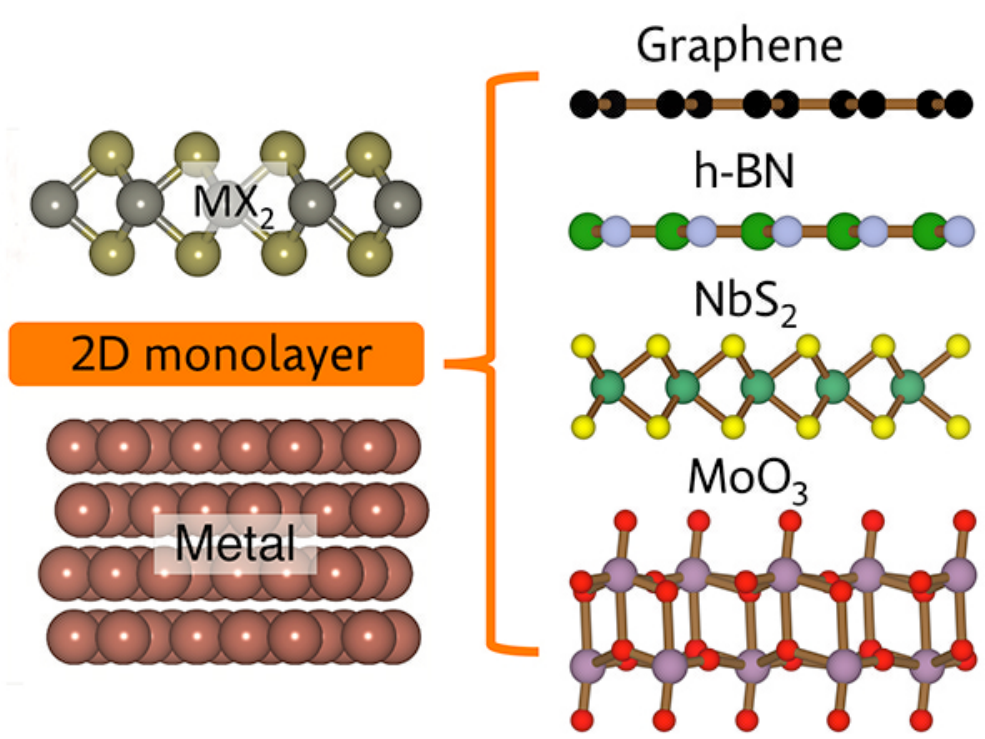}
\caption{Side view of the metal/buffer/\MX\ structure, M = Mo,W, X = S,Se,Te with possible buffer layers graphene, \BN, \NbS, and \MO. Visualization by VESTA.\cite{Momma:db5098} }
\label{Fig1}
\end{figure}

We suggest a practical way to solve the p-type contact problem and tune the SB height by inserting a monolayer of a 2D material between the metal substrate and the \MX\ semiconductor, see figure~\ref{Fig1}. The buffer layer suppresses the metal/\MX\ interface states. 2D materials have certain unique properties not found in buffer layers of 3D materials.\cite{Geim:nat13} As the interlayer bonding is van der Waals, neither the structure of the 2D buffer layer, nor that of the \MX\ semiconductor, is perturbed significantly by stacking them. The 2D buffer layer need not be lattice matched to the metal or to the \MX\ layer, and the structure of the multilayer will in general be incommensurate. Van der Waals interface bonding also promises the absence of interface states.  Covering the metal by an adsobant layer such as graphene, a monolayer of hexagonal boron nitride (\BN) or $T$-\MS, has proved to be beneficial for making n-type contacts.\cite{Farmanbar:prb15,Chhowalla:natmat14} We show that a 2D buffer layer can be selected to obtain a zero SB height for holes.

\begin{figure*}[t]
\includegraphics[width=14cm, angle=0]{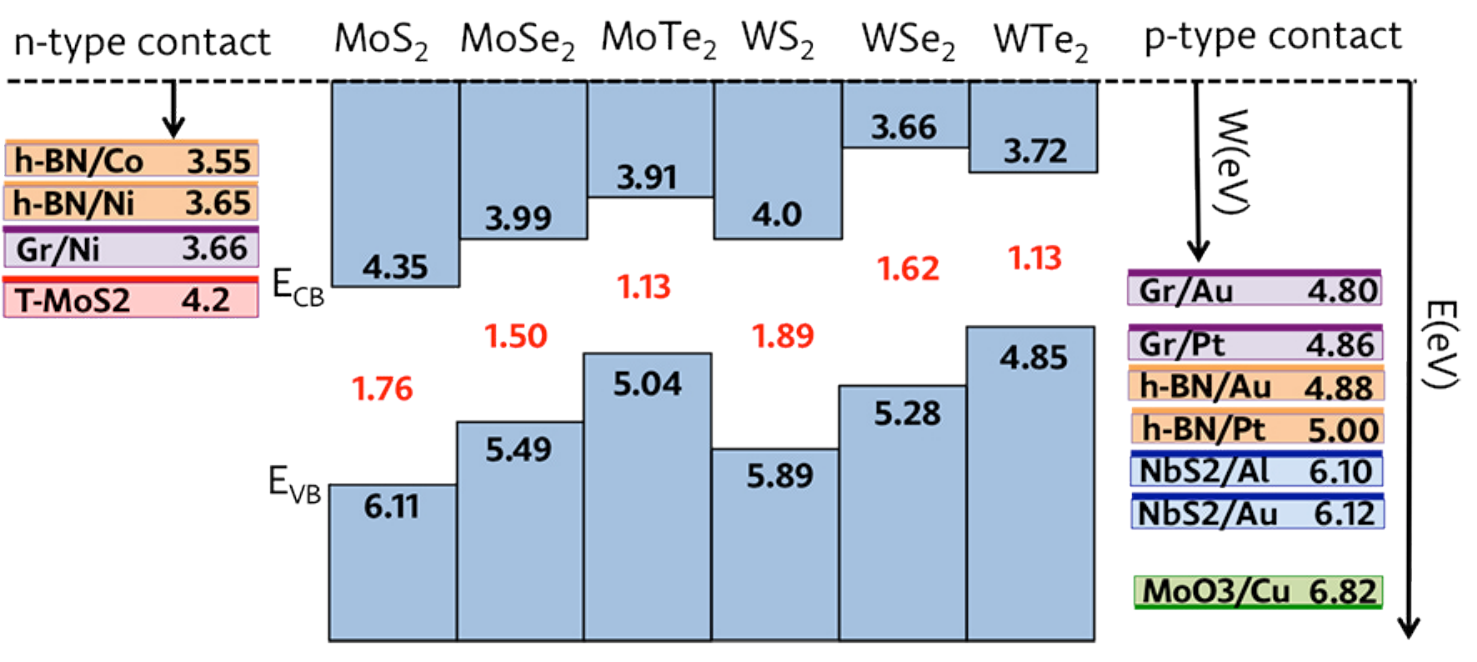}
\caption{Middle: calculated valence band maxima $E_\mathrm{VB}$ and conduction band minima $E_\mathrm{CB}$ of $H$-\MX\ monolayers (band gaps are given in red), see ``Computational section''. Left: work functions of n-type metal/buffer contacts.\cite{Chhowalla:natmat14,Giovannetti:prl08,Khomyakov:prb11,Bokdam:prb14a,Farmanbar:prb15} Right: calculated work functions of p-type metal/buffer contacts.}
\label{Fig2}
\end{figure*}

A \BN\ monolayer is a buffer layer that can be used for making n-type contacts, because adsorption of \BN\ on a metal decreases its work function by up to 1-2 eV.\cite{Bokdam:prb14a} For instance, Co/\BN\ and Ni/\BN\ are predicted to form zero SB height n-type contacts to \MX\ semiconductors, see figure~\ref{Fig2}.\cite{Farmanbar:prb15} A decrease of the work function is unfavorable for making p-type contacts. Metal/\BN\ gives a p-type contact to \MX\ only if the metal work function is sufficiently high, and the \MX\ ionization potential is sufficiently low. We find a zero SB height for Pt/\BN, and Au/\BN\ contacts to \MTe\ and a low SB for Pt/\BN/\WSe, see figure~\ref{Fig2}. Alternatively, one can use a graphene buffer layer,\cite{Chuang:nanol14,Leong:acs15} whose behavior is qualitatively similar to that of a \BN\ monolayer. 

To make more universely applicable p-type contacts, one needs a buffer layer that effectively increases the metal work function. Oxides such as \MO\ are an option. The \MO\ structure consists of bilayers, making it conceivable to cover a metal with a single \MO\ bilayer, see figure~\ref{Fig1}. We find that the electron affinity of \MO\ is sufficiently high to make carrier transport through its conduction band possible, so that a bilayer of \MO\ does not present a tunnel barrier. 

A very interesting option for creating p-type contacts is to use metallic \MXP\ buffer layers, such as \NbS\ or \TaS.\cite{Mattheis:prb73} Their structure is similar to that of semiconducting \MX, they are chemically stable, and have work functions close to 6 eV. We show that a monolayer \NbS\ adsorbed on a metal gives a SB with zero height for contacts to all \MX. The initital work function of the metal substrate is irrelevant; Au/\NbS\ and Al/\NbS\ essentially give the same contact.

\section{Results and Discussion}

\subsection{Van der Waals bonded contacts}
The obvious way to make a p-type contact to a semiconductor is to use a metal with a high work function. The calculated work function of Pt is 5.91 eV, suggesting that this metal should give a zero SB to all \MX, except \MS. In practice this is not true, as \MX\ interacts with Pt to give states at the interface whose energy is within the band gap of \MX. Initially it was thought that \MX\ could escape the formation of interface gap states (IGS), as, unlike conventional semiconductors such as Si, \MX\ has no dangling surface bonds that interact strongly with the metal surface.\cite{Lince:prb87} Nevertheless, even a relatively weak interaction yields IGS that pin the Fermi level in the gap, which results in an appreciable SB.\cite{Farmanbar:prb15} 

\begin{figure}[t]
\includegraphics[width=12cm, angle=0]{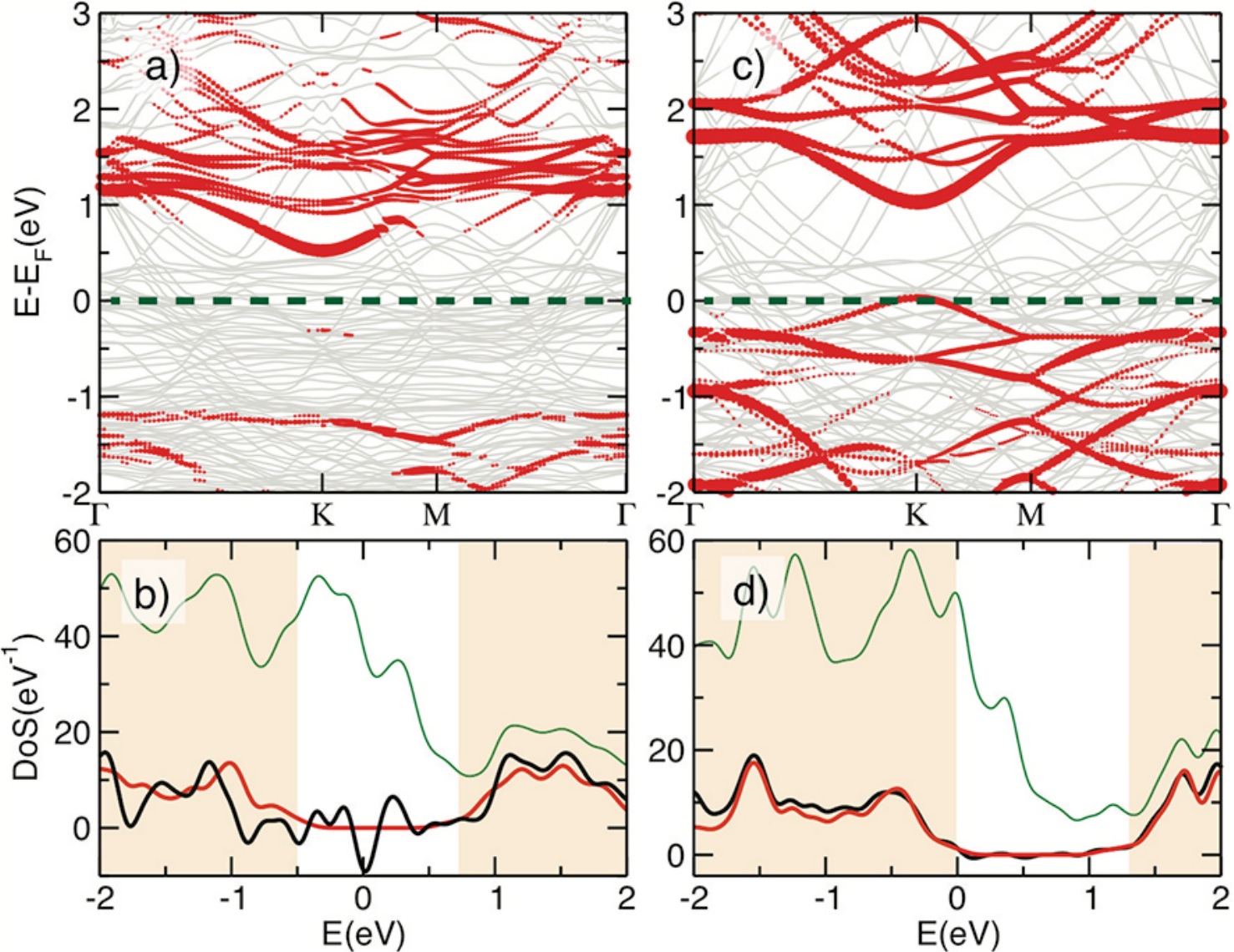}
\caption{(a) Band structure (grey) of the Pt(111)/\MTe\ slab; the bands are colored red according to the size of a projection of the corresponding wave functions on the \MTe\ atoms. The zero of energy is set at the Fermi level. (b) Green: total density of states, DoS$_\mathrm{tot}$, of the Pt(111)/\MTe\ slab; black: DoS$_\mathrm{tot}$ minus the DoS of the clean Pt(111) slab, $\Delta\mathrm{DoS} = \mathrm{DoS}_\mathrm{tot} - \mathrm{DoS}_\mathrm{Pt}$; red: the DoS of a free-standing \MTe\ layer, DoS$_\mathrm{MoTe_2}$.  
The \MTe\ band gap region is indicated in white. (c,d) Band structure and DoSs of the Pt(111)/\BN/\MTe\ system.}
\label{Fig3}
\end{figure}

As an example, figure~\ref{Fig3}(a) gives the band structure of the Pt(111)/\MTe\ interface. The direct interaction between \MTe\ and the Pt surface perturbs the band structure of \MTe\ significantly, the valence bands in particular. The perturbation is accompanied by the formation of IGS inside the \MTe\ band gap, figure~\ref{Fig3}(b), which pin the Fermi level. The SB height is defined as 
\begin{equation}
\Phi_\mathrm{p} = E_\mathrm{VB} - E_\mathrm{F},
\end{equation}
with $E_\mathrm{F}$ the Fermi energy, and $E_\mathrm{VB}$ the energy of the top of the valence band (measured as distances, i.e., positive numbers, from the vacuum level). The SB to an electronically perturbed overlayer is of course not extremely well-defined. One can estimate the SB by aligning the core levels of the adsorbed \MTe\ layer to those of a free-standing \MTe\ layer, which gives $\Phi_\mathrm{p} = 0.49$ eV. With $E_\mathrm{VB}=5.04$ eV (figure~\ref{Fig1}) this gives $W_{\mathrm{M}|\mathrm{WTe}_2}=4.55$ eV as the work function of of Pt covered by a \MTe\ monolayer. As the calculated work function of clean Pt(111) is 5.91 eV, it implies that adsorbing \MTe\ on Pt creates a large potential step at the interface of 1.36 eV.

The changes in the electronic structure of the adsorbed \MTe\ layer are visualized in figure~\ref{Fig3}(b). Starting from the total density of states (DoS) of the Pt(111)/\MTe\ system, and subtracting the DoS of the clean Pt(111) slab, one can compare the result to the DoS of a free-standing \MTe\ layer. The comparison shows considerable differences in the \MTe\ band gap region, which are direct evidence for the formation of IGS.  

The pattern of these IGS depends on the particular metal/\MX\ combination. We have however not found an elemental high work function metal that does not give IGS. For all high work function metal/\MX\ contacts we have studied, IGS are formed that pin the Fermi level and yield a sizable SB. The same problem has emerged previously for low work function metals and n-type SBs. Introducing a buffer layer can break the interaction between \MX\ and the metal and eliminate the IGS. This layer must be sufficiently thin, such that it does not form a large barrier for the charge carriers. In addition, the interaction between \MX\ and the buffer layer must not create new IGS. 

A single atomic layer of graphene or \BN\ obeys these criteria. Such a layer presents only a thin barrier that essentially allows for metallic transport through the layer.\cite{Yazyev:prb09} Inserting graphene or a \BN\ monolayer between a metal surface and \MX\ disrupts the metal-\MX\ chemical interaction, and destroys any metal-induced IGS. Graphene or \BN\ bind to \MX\ via van der Waals forces. One does not expect such an interaction to create new IGS. This is illustrated for Pt(111)/\BN/\MTe\ in figures~\ref{Fig3}(c) and (d). Inserting a monolayer of \BN\ restores the electronic structure of \WTe, where the projected bands are essentially those of free-standing \MTe. The DoS of Pt(111)/\BN/\MTe\ minus that of Pt(111) is essentially identical to the DoS of a free-standing \MTe\ layer, in particular in the gap region. In other words, it shows no sign of IGS generated by any \BN/\MTe\ interaction.  

The concept also works if one uses graphene to cover the metal. The interaction between metal/graphene or metal/\BN\ and \MX\ is van der Waals, so the electronic structure of any \MX\ is preserved, and IGS are absent. A serious drawback however is that adsorption of graphene or \BN\ generally decreases the metal work function considerably, e.g., by 0.6-1.1 eV for Pt and Au, see table~\ref{tab:graphene}. The reduction originates from a dipole layer that is formed at the metal/graphene or metal/h-BN interface, where Pauli exchange repulsion gives an dominant contribution.\cite{Bokdam:prb14b} 

\begin{figure}[t]
\includegraphics[width=7cm]{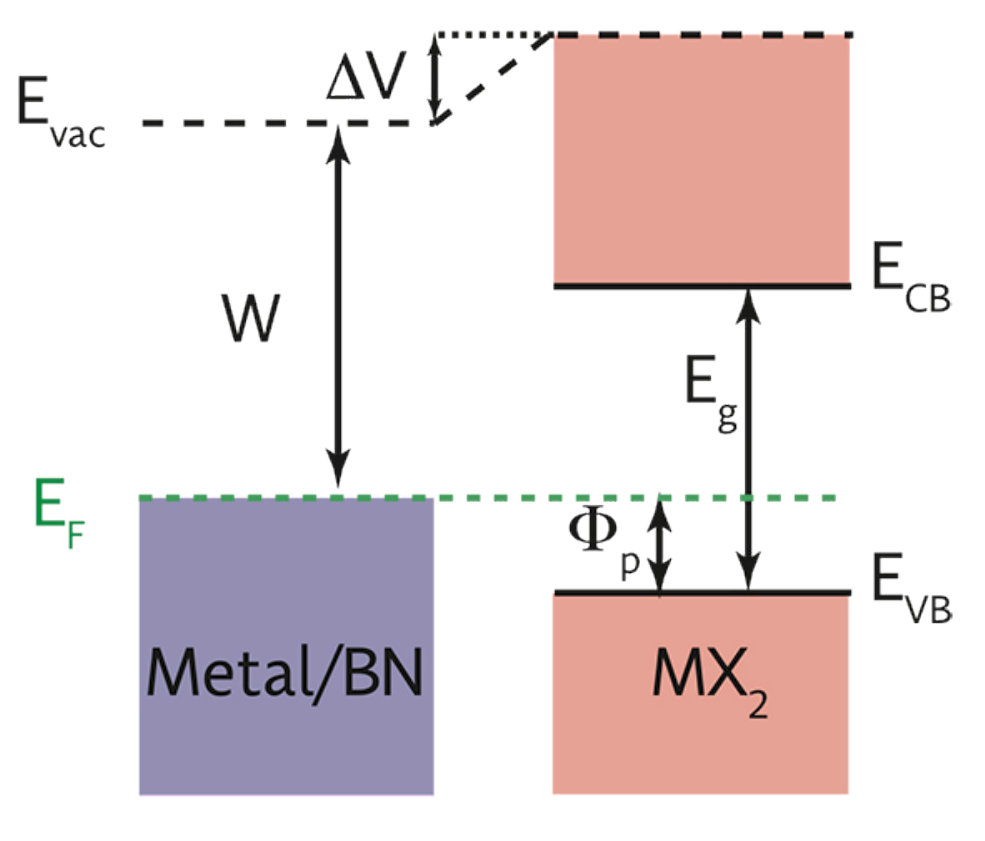}
\caption{Schematic diagram defining the energy parameters used to describe the interface region in metal/\BN/\MX\ structures, with E$_\mathrm{vac}$ and E$_\mathrm{F}$ the vacuum level and the Fermi level, $W$ and $\Delta V$ the work function of metal/\BN\ and the potential step at the \BN/\MX\ interface, E$_\mathrm{VB}$ and E$_\mathrm{CB}$, the top of the \MX\ valence band and the bottom of the conduction band, $E_g$ the band gap, and $\Phi_\mathrm{p}$ the SB height for holes.}
\label{Fig4}
\end{figure}

The reduction is partly canceled by a potential step $\Delta V$ formed at the graphene/\MX\ or \BN/\MX\ interface, see figure~\ref{Fig4}. Although the weak interaction between \BN\ and \MX\ does not give IGS, it does lead to an interface potential step, which originates from the Pauli repulsion between the electrons from the \BN\ layer and those originating from the \MX\ layer.\cite{Bokdam:prb14b} As the chalcogenide atoms form the outer atomic layers in \MX, it is then not surprising that $\Delta V$ depends on the chalcogenide species, but not on the metal species of the central atomic layer, see table~\ref{tab:graphene}. For graphene/\MX\ and \BN/\MX\ interfaces $\Delta V$ is positive toward \MX\, i.e., it decrease the SB. The steps are actually quite moderate, i.e. in the range 0.2-0.3 eV, see table~\ref{tab:graphene}, so they do not cancel out the work function reductions discussed in the previous paragraph. 

\begin{table}[t]
\caption{Calculated work functions $W$ of metal/graphene, metal/\BN\ and metal/\MS\ substrates; potential steps $\Delta V$ at graphene/\MX, \BN/\MX\ and \MS/\MX\ interfaces;  SB heights $\Phi_\mathrm{p}$ (eq.~\ref{eq1}). All values in eV; all \MX\ in the $H$-structure. The calculated work functions of clean Pt(111) and Au(111) surfaces are 6.04 and 5.58 eV, respectively.}
%\begin{ruledtabular}
  \begin{tabular}{lccccccccc}
       &                  & \multicolumn{2}{c}{\MSe} & \multicolumn{2}{c}{\WSe}  & \multicolumn{2}{c}{\MTe} & \multicolumn{2}{c}{\WTe} \\
       \cline{3-4} \cline {5-6} \cline{7-8} \cline {9-10}
       & $W$     &   $\Delta V$    & $\Phi_\mathrm{p}$ &   $\Delta V$    & $\Phi_\mathrm{p}$ &   $\Delta V$    & $\Phi_\mathrm{p}$ &   $\Delta V$    & $\Phi_\mathrm{p}$ \\ 
       \hline
Pt/Gr      &   4.86  &  0.21   & 0.42   &  0.21   & 0.21   &  0.28   & 0 &  0.28   & 0  \\
Pt/\BN\  &   5.00  &  0.18    &  0.31 &  0.18    &  0.10  &  0.24  & 0 &  0.24  & 0 \\
Pt/\MS\ &   5.32  &  0.14    &  0.03  &  0.14   &  0        &  0.18  & 0 &  0.18  & 0 \\
Au/Gr    &    4.80  &  0.21   &  0.48  &  0.21   &  0.27   &  0.28  & 0 &  0.28  & 0 \\
Au/\BN\ &   4.88  &  0.18   &  0.43   &  0.18   &  0.22   &  0.24  & 0 &  0.24  & 0 \\
Au/\MS\ &   5.05  &  0.14    &  0.30  &  0.14   &  0.09   &  0.18  & 0 &  0.18  & 0 \\
\hline
\end{tabular}
%\end{ruledtabular}
\label{tab:graphene}
\end{table}   

One obtains a zero SB height only with graphene- or \BN-covered metals with a high work function, such as Pt and Au, in contact with \MX\ that has a sufficiently low ionization potential. The SB height can be calculated from the numbers given in table~\ref{tab:graphene} and figure~\ref{Fig1}
\begin{equation}
\Phi_\mathrm{p} = E_\mathrm{VB} - W - \Delta V.
\label{eq1}
\end{equation}
As by definition $\Phi_\mathrm{p}\geq 0$, a negative number indicates a zero SB height, $\Phi_\mathrm{p}=0$. Only the tellurides \MTe\ and \WTe\ satisfy this criterion. \WSe\ gives a small SB of 0.10 eV with Pt/\BN\, but the other selenide monolayers give appreciable SB heights in the range 0.2-0.5 eV. The sulfides (not shown in table~\ref{tab:graphene}) have large SBs with heights $\sim 1$ eV.

One can however expect that the situation becomes more favorable for \MX\ multilayers as the band gap of multilayer \MX\ is smaller than that of a \MX\ monolayer. For \MS\ it has been argued that the band gap reduction is monotonic in the number of layers and that it is equally divided into an upward shift of the valence band and a downward shift of the conduction band.\cite{Song:ACS8} This is likely to hold more generally for all \MX\ compounds. Indeed an explicit calculation of bilayer \WSe\ on Pt/\BN\ gives a SB that is zero. It implies that vanishing SBs to multilayer \WSe\ are possible with graphene- or \BN-covered high work function metals.

Adding additional \BN\ layers on top of a \BN/Au or \BN/Pt substrate does not change the SB height, as the potential steps formed at the interfaces between the \BN\ layers are negigibly small. By itself \BN\ is an insulator forming SBs for holes with heights 0.9-1.1 eV and 1.1-1.3 eV with Pt and Au, respectively.\cite{Bokdam:prb14a} Therefore, \BN\ acts as a tunnel barrier between Pt or Au and \MX. A single \BN\ layer forms only a thin barrier that is very transparent,\cite{Yazyev:prb09} but the contact resistance is expected to grow exponentionally with the number of \BN\ layers.\cite{Britnell:nanol12}

The principles outlined above can be extended to other buffer layers besides graphene or \BN. For instance, adsorbing a single \MS\ layer on a high work function metal such as Au or Pt reduces its work function by 0.5-0.7 eV. That still leaves us with a substrate that has a sufficiently high work function to yield p-type contacts to the tellurides with a zero SB height, and a small to zero SB height to the selenides, see table~\ref{tab:graphene}. As the Fermi level for \MS\ adsorbed on Au or Pt is close to the middle of the \MS\ band gap,\cite{Farmanbar:prb15} the \MS\ layer then acts as tunnel barrier between the metal and the \MX\ layer. Because a \MS\ monolayer is thicker than graphene or \BN, one expects it to present more of a barrier, and yield a higher contact resistance.

\subsection{High electron affinity oxide layers}
A buffer layer that effectively decreases a metal work function limits the scope for using it to create p-type contacts. A buffer layer that increases the work function would be more advantageous, which requires a layer that accepts electrons from the underlying metal. Oxides such as \MO\ and \WO\ are known for their potential as p-type contacts in organic photovoltaics and light-emitting diodes,\cite{Greiner:afm13,Kroeger:apl09} and have also been tested in field-effect transistors based upn \MX.\cite{Chuang:nanol13,McDonnell:acs8} More specifically, $\alpha$-\MO\ (the thermodynamically stable phase of \MO) is a layered material, which consists of covalently bonded bilayers, see 􏰇figure~\ref{Fig1}, that are van der Waals stacked. \MO\ is a large band gap material with an experimental band gap of 3.0 eV. The electron affinity of this material is however an exceptionally high 6.7-6.9 eV.\cite{Greiner:afm13,Kroeger:apl09} Therefore, \MO\ is predicted to be an electron acceptor with respect to common metals.

\begin{figure}[t]
\includegraphics[width=8cm]{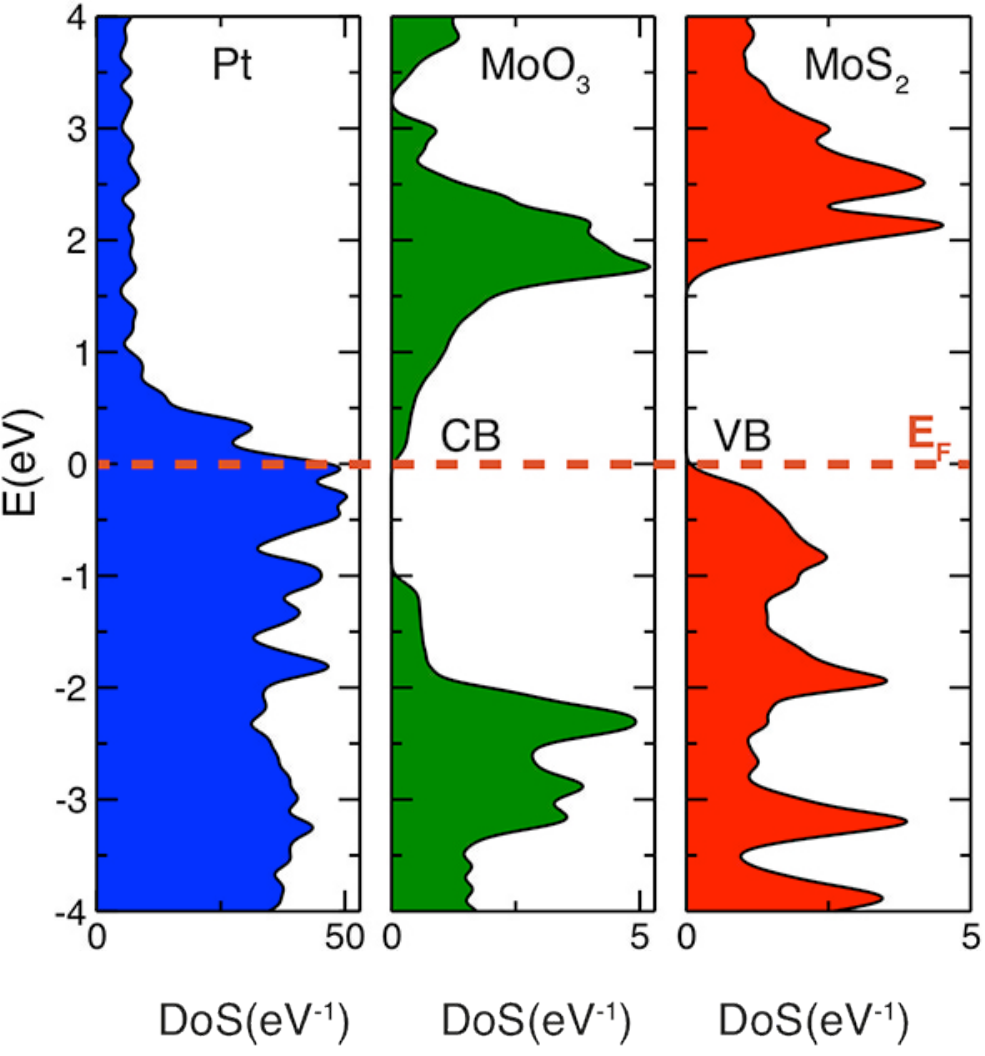}
\caption{Densities of states of Pt(111) (left), bilayer-\MO\ (middle), and \MS\ (right). In a Pt/bilayer-\MO/\MS\ multilayer the Fermi level $E_\mathrm{F}$ is pinned at the bottom of the \MO\ conduction band $E_\mathrm{CB}$ and at the top of the \MS\ valence band $E_\mathrm{VB}$.}
\label{Fig5}
\end{figure}

Adsorbing a \MO-bilayer on a common metal leads to a transfer of electrons from the metal to the \MO, and sets up a dipole layer that effectively increases the work function, provided that the adsorption process does not destroy the structure and integrity of the \MO\ overlayer. For instance, the calculated electron work function of Cu(111)/bilayer-\MO\ is 7.08 eV, as compared to 4.98 eV for the clean Cu(111) surface. Moreover, the density of states of the \MO\ conduction band is sufficiently high such as to pin the Fermi level at the bottom of the conduction band, see figure~\ref{Fig5}.  

A \MO\ bilayer has no dangling bonds. A \MX\ layer adsorbed on \MO\ is therefore likely to be van der Waals-bonded. Indeed from our calculations we find a small \MO-\MS\ equilibrium binding energy of $0.17$ eV per \MS\ formula unit, and a large equilibrium bonding distance (between to top layer of oxygen atoms and the bottom layer of sulfur atoms) of 3.1 \AA. The weak \MS/\MO\ bonding has little influence on the electronic structure of either layer. Any metal that can be covered by bilayer-\MO\ without disrupting its structure should then give a work function $W$ that is sufficiently high to give a zero SB height to all \MX, see figure~\ref{Fig2}. As $W>E_\mathrm{VB}$ electrons are transferred from \MX\ to \MO, thereby pinning the Fermi level at the top of the \MX\ valence band, as well as at the bottom of the \MO\ conduction band, see figure~\ref{Fig5}. 

The \MO\ layer does not act as a tunnel barrier, as transport of charge carriers takes place through the conduction band of the oxide. If one adds additional \MO\ layers, one expects that for thicknesses below the mean free path of the charge carriers the characteristics for injection from the metal into the \MX\ layer remain the same. The contact resistance does increase for thicker \MO\ layers, however, as charge carrier mobilities in oxide layers are typically substantially smaller than in metals.\cite{Chuang:nanol13}

In principle any oxide that has a layered structure can be used this way. One can cover a metal with an oxide monolayer; if the oxide layer bonds to  \MX\ through van der Waals interactions, and if its electron affinity is sufficiently high, it should give a p-type contact. 
In contrast to graphene or \BN, the metal species is not very important, as the conduction band of the oxide pins the Fermi level. Besides \MO, for instance V$_{2}$O$_{5}$ has a layered structure and a high electron affinity of up to 7 eV.\cite{Greiner:afm13} A drawback of using such oxide layers are that they are strong oxidizing agents, which can react with molecules in the environment, or with the metal electrodes. For instance, the Cu/\MO\ interface is metastable and oxidized Cu ions can diffuse into \MO.\cite{Greiner:afm13} Other metal/\MO\ interfaces, such as Au/\MO, are  stable, however.

\subsection{Metallic \MXP\ buffer layers}

A buffer with a high electron affinity that is less reactive than an oxide would be very convenient. Metallic \MXP, $\mathrm{M}'=\mathrm{V,Nb,Ta}$, $\mathrm{X}'=\mathrm{S,Se}$ compounds have a layered hexagonal structure similar to that of semiconducting \MX. The latter compounds contain a group VI transition metal M, whereas the former compounds contain a group V transition metal $\mathrm{M}'$.
The electronic structure of these two compound groups is basically similar, but in \MX\ the topmost valence band is completely filled, whereas in \MXP\ it is only half-filled because $\mathrm{M}'$ has one electron less than M.\cite{Mattheis:prb73} This makes the \MXP\ compounds metallic with a relatively high work function. For example, \NbS\ and \TaS\ monolayers have calculated work functions of 6.22 and 6.12, respectively.

Most common metals have a lower work function. If they are covered by a \MXP\ monolayer, electrons are transferred from the metal to the \MXP\ layer, effectively increasing the work function, as in the case of an \MO\ overlayer. As the density of states at the Fermi level of \MXP\ is high, the transferred electrons will hardly modify the Fermi level, which is therefore effectively pinned by the \MXP\ layer. For instance, the Au(111) and Al(111) surfaces have calculated work functions of 5.43 and 4.24 eV, respectively. Adsorbing a \NbS\ monolayer gives work functions of Au/\NbS\ and Al/\NbS\ surfaces of 6.11 and 6.06 eV, respectively. In other words, the work function differs by $\lesssim 0.15$ eV from that of a free-standing \NbS\ monolayer, irrespective of the metal substrate.

This remarkable property make metallic \MXP\ compounds viable buffer layers for making p-doped contacts to \MX\ semiconductors. We envision that \MXP\ layers can be deposited on a metal substrate in a similar way as \MX\ layers. A \MX\ layer can then be deposited on \MXP\ by van der Waals epitaxy, for instance.\cite{Lee:advmat12,Chen:advmat15} The interaction at the \MXP/\MX\ interface is van der Waals, which hardly perturbs the electronic structure of either layer. The potential step at the \MXP/\MX\ interface, $\Delta V\approx 0.1$ eV, is small and hardly affects the \MXP\ work function.  

\begin{figure}[t]
\includegraphics[width=9cm, angle=0]{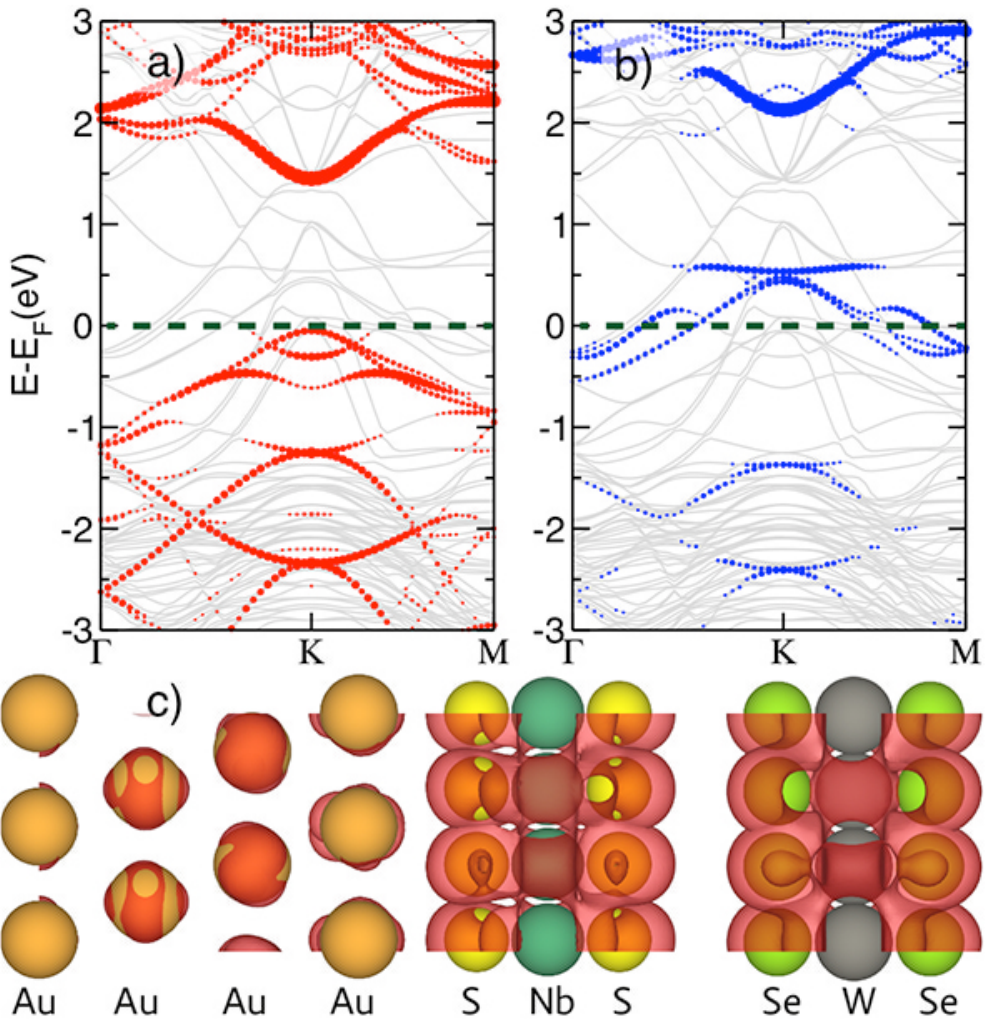}
\caption{(a,b) Band structure of Au(111)/\NbS/\WSe. The red and blue colors indicate projections of the wave functions on the \WSe\ and the \NbS\ atoms, respectively. The zero of energy is set at the Fermi level. (c) Local density of states calculated in an energy interval of 0.1 eV around the Fermi level.}
\label{Fig6}
\end{figure}

Figure~\ref{Fig6} shows the calculated band structure of the Au(111)/\NbS/\WSe\ multilayer. It illustrates the perfect p-type contact formed in this case, with the Fermi level coinciding with the top of the \WSe\ valence band. At the same time the Fermi cuts the valence band of \NbS, confirming that it acts as a conducting layer. Indeed, the local DoS calculated at the Fermi energy shows a state that is delocalized over the whole multilayer, see figure~\ref{Fig6}. We expect that similar p-type contacts can be formed with other \MX\ layers using \NbS\ or \TaS\ monolayers as a buffer.

Adding additional \NbS\ or \TaS\ layers, the charge carrier injection into \MX\ should remain the same as long as the thickness of the buffer layers is below the mean free path of the charge carriers. For thicker layers the contact resistance starts to depend on the thickness of the buffer layers.

%************************************************CONCLUSION*****************************************
\section{Conclusions}
We propose a particular technique to construct p-type contacts with zero SB heights to \MX\ (M = Mo,W; X = S,Se,Te) 2D semiconductors. Using first-principles DFT calculations we show that a direct metal/\MX\ interaction leads to interface states that pin the Fermi level in the \MX\ band gap, giving a sizable SB. Inserting a well-chosen buffer layer between the metal surface and the \MX\ layer breaks this interaction, and unpins the Fermi level, if \MX\ is bonded to the buffer layer with van der Waals forces.   

A monolayer of \BN\ or graphene constitute only a thin barrier for transport. Adsorbing \BN\ on a high work function metal such as Pt or Au, gives a zero SB height for contacts to the \MX\ tellurides, a low SB to \WSe, and zero SB heights for contacts to telluride and selenide multilayers. Adsorbing graphene works in a similar way, but is somewhat less effective in reducing SBs. To obtain p-type contacts one has to combine these layers with high work function metals, as adsorption of \BN\ or graphene substantially reduces the work function of the metal substrate. 

Alternatively one can use a buffer layer of an oxide with a high electron affinity. By explit calculations on metal/\MO\ we show that adsorbing a bilayer of \MO\ gives a zero SB height for contacts to all \MX. The metal substrate is relatively unimportant here, as the Fermi level is pinned at the \MO\ conduction band, which also ensures  that bilayer-\MO\ does not act as a tunnel barrier to \MX. Bilayer-\MO\ binds to \MX\ layers via van der Waals interactions, consistent with the principal idea of this paper. 

Strong oxididants such as \MO\ might suffer from chemical instability. The most elegant solution for the p-type contact problem is using a metallic \MXP\ (M = Nb,Ta; X = S) monolayer as a buffer. These compounds are stable and have a high work function. Adsorbed on a metal substrate they yield a high work function contact, which gives a zero SB height to all \MX\ monolayers. The metal substrate is of little importance, as the Fermi level is pinned by the \MXP\ layer. \MX\ forms a van der Waals bonded contact to \MXP, and the states formed at the Fermi level extend throughout the whole metal/\MXP/\MX\ multilayer.

%*****************computational
\section{Computational section}
We perform first-principles density functional theory (DFT) calculations using the VASP code and the projector augmented wave (PAW) datasets from the VASP database.\cite{Kresse:prb93,Kresse:prb96,Blochl:prb94b,Kresse:prb99} For the computational set-up and the choice of computational parameters (unit cell, $k$-point sampling, etc.) we follow Refs.~\cite{Farmanbar:prb15} and \cite{Farmanbar:arxiv}. 

Obviously different 2D materials have different in-plane lattice constants. Van der Waals stacking of 2D materials then generally leads to incommensurable structures, which are observed experimentally in the form of moir\'{e} patterns. In electronic structure calculations one is forced to approximate these by commensurable structures, which introduces an artificial strain. As the \MX\ electronic structure is very sensitive to strain, we keep the in-plane lattice constant of \MX\ at its optimized value, adapting the lattice constant of the metal surfaces and the buffer layers accordingly. The supercells defining the commensurable structures are constructed such that the lattice mismatches between the different layers are minimal, following the technique explained in Ref.~\cite{Farmanbar:arxiv}. As an example, a (2$\times$2) \MTe\ cell on top of (3$\times$3) \BN\ on ($\sqrt{7}$$\times$$\sqrt{7}$) Pt gives lattice mismatches of 6 and 4\% in the \BN\ and Pt lattices respectively.

\begin{table}[t]
\caption{Optimized in-plane lattice constant $a$(\AA) and calculated band gap $E_g$(eV) of the $H$-\MX\ monolayers. All calculated values are obtained using opt86b-vdw-DF.}
%\begin{ruledtabular}
\begin{tabular}{llllllll}
         &\MS\ & \MSe\ & \MTe\ & \WS\ & \WSe\ & \WTe\ & \NbS\ \\
\hline
$a$                                                  & 3.16 & 3.29 & 3.52 & 3.16 & 3.29 & 3.52 & 3.31 \\
$a_\mathrm{exp}^\mathrm{a)}$ & 3.16 & 3.30 & 3.52 & 3.15 & 3.28 &  ---  & 3.31 \\
$E_g$                                          & 1.76 & 1.50  & 1.13 & 1.89 & 1.62 & 1.13 & --- \\
$E_\mathrm{opt}^\mathrm{exp}$  & 1.86$^\mathrm{b),c),d)}$ & 1.63$^\mathrm{d),e)}$ & 1.10$^\mathrm{f)}$ & 1.99$^\mathrm{g)}$ & 1.65$^\mathrm{b)}$ & --- & --- \\
\hline
\end{tabular}
%\end{ruledtabular}
 \\$^\mathrm{a)}$ Refs.~\cite{Boker:prb01,Schutte:jssc87,Ohno:prb98}, $^\mathrm{b)}$ Ref.~\cite{Chiu:natcom15}, $^\mathrm{c)}$, Ref.~\cite{Mak:prl10} $^\mathrm{d)}$ Ref.~\cite{Tongay:nanolett12}, $^\mathrm{e)}$ Ref.~\cite{Ugeda:natmat14}, $^\mathrm{f)}$ Ref.~\cite{Ruppert:nanolett12}, $^\mathrm{g)}$ Ref.~\cite{Gutierrez:nanolett13}.  
\label{tab:lat}
\end{table}

Van der Waals interactions are vital to descibe the bonding between the layers. Here we use the van der Waals density functional opt86b-vdW-DF.\cite{Dion:prl04,Klimes:prb11,Thonhauser:prb76} The optimized lattice parameters $a$ of \MX\ monolayers are in good agreement with the experimental lattice parameters of the bulk hexagonal structures, see table~\ref{tab:lat}. (The experimental structure  of \WTe\ is not hexagonal;\cite{Dawson:jpc87} it has been added for completeness reasons.) Also in table~\ref{tab:lat} are listed the calculated monolayer band gaps $E_g$. DFT calculations usually severely underestimate band gaps, but for \MX\ monolayers the differences are tolerable, if one compares to the measured optical gaps $E_\mathrm{opt}^\mathrm{exp}$ of the monolayers, see table~\ref{tab:lat}. 

Such a comparison is not entirely fair, as $E_g-E_\mathrm{opt}^\mathrm{exp}$ is the exciton binding energy. The latter depends  strongly on the environment of the exciton, and decreases if the electrodynamic screening by the environment increases.\cite{vanderhorst:prl99} Values of the exciton binding energy between zero,\cite{Mak:prl10,Zhang:natnano14} and 0.5 eV\cite{Chiu:natcom15,Ugeda:natmat14} have been reported. In our case one expects the screening to be strong as the \MX\ layers are close to a metal substrate, and we speculate the exciton binding energy to be relatively small. 

Comparing to previous calculations, one observes that the opt86b-vdw-DF functional gives an improvement over the GGA/PBE functional regarding optimized lattice constants and band gaps of \MX.\cite{Kang:apl13} The opt86b-vdw-DF lattice constants are close to the HSE06 values, and the band gaps are in between the GGA/PBE and HSE06 values. Concerning the absolute positions of the energy levels, the valence band maxima $E_\mathrm{VB}$ calculated with opt86b-vdw-DF (figure~\ref{Fig1}) and HSE06\cite{Kang:apl13} are generally with $\sim0.15$ eV of one another. It is difficult to compare these values to experimental data. Note however that the experimentally determined valence band offset of 0.83 eV between \MS\ and \WSe,\cite{Chiu:natcom15} is in excellent agreement with the 0.83 eV derived from the numbers in figure~\ref{Fig1}.

%************************************************Metal\BN\TMD*****************************************

\section{Acknowledgment}
This work is part of the research program of the Foundation for Fundamental Research on Matter (FOM), which is part of the Netherlands Organisation for Scientific Research (NWO). The use of supercomputer facilities was sponsored by the Physical Sciences Division (EW) of NWO.

%\begin{thebibliography}{99}

\providecommand{\latin}[1]{#1}
\providecommand*\mcitethebibliography{\thebibliography}
\csname @ifundefined\endcsname{endmcitethebibliography}
  {\let\endmcitethebibliography\endthebibliography}{}

%\end{thebibliography}


\begin{mcitethebibliography}{58}
\providecommand*\natexlab[1]{#1}
\providecommand*\mciteSetBstSublistMode[1]{}
\providecommand*\mciteSetBstMaxWidthForm[2]{}
\providecommand*\mciteBstWouldAddEndPuncttrue
  {\def\EndOfBibitem{\unskip.}}
\providecommand*\mciteBstWouldAddEndPunctfalse
  {\let\EndOfBibitem\relax}
\providecommand*\mciteSetBstMidEndSepPunct[3]{}
\providecommand*\mciteSetBstSublistLabelBeginEnd[3]{}
\providecommand*\EndOfBibitem{}
\mciteSetBstSublistMode{f}
\mciteSetBstMaxWidthForm{subitem}{(\alph{mcitesubitemcount})}
\mciteSetBstSublistLabelBeginEnd
  {\mcitemaxwidthsubitemform\space}
  {\relax}
  {\relax}

\bibitem[Chhowalla \latin{et~al.}(2013)Chhowalla, Shin, Eda, Li, Loh, and
  Zhang]{Chhowalla:natchem13}
Chhowalla,~M.; Shin,~H.~S.; Eda,~G.; Li,~L.-J.; Loh,~K.~P.; Zhang,~H.
  \emph{Nature Chemistry} \textbf{2013}, \emph{5}, 263Ð275\relax
\mciteBstWouldAddEndPuncttrue
\mciteSetBstMidEndSepPunct{\mcitedefaultmidpunct}
{\mcitedefaultendpunct}{\mcitedefaultseppunct}\relax
\EndOfBibitem
\bibitem[Cui \latin{et~al.}(2015)Cui, Xin, Yu, Pan, Ong, Wei, Wang, Nan, Ni,
  Wu, Chen, Shi, Wang, Zhang, Zhang, and Wang]{Cui:advmat15}
Cui,~Y. \latin{et~al.}  \emph{Adv. Mater.} \textbf{2015}, \emph{27},
  5230--5234\relax
\mciteBstWouldAddEndPuncttrue
\mciteSetBstMidEndSepPunct{\mcitedefaultmidpunct}
{\mcitedefaultendpunct}{\mcitedefaultseppunct}\relax
\EndOfBibitem
\bibitem[Wang \latin{et~al.}(2012)Wang, Kalantar-Zadeh, Kis, Coleman, and
  Strano]{Wang:nnano12}
Wang,~Q.~H.; Kalantar-Zadeh,~K.; Kis,~A.; Coleman,~J.~N.; Strano,~M.~S.
  \emph{Nat. Nanotechnol.} \textbf{2012}, \emph{7}, 699--712\relax
\mciteBstWouldAddEndPuncttrue
\mciteSetBstMidEndSepPunct{\mcitedefaultmidpunct}
{\mcitedefaultendpunct}{\mcitedefaultseppunct}\relax
\EndOfBibitem
\bibitem[Zhang \latin{et~al.}(2013)Zhang, Huang, Chen, Chang, Cheng, and
  Li]{Zhang:advmat13}
Zhang,~W.; Huang,~J.-K.; Chen,~C.-H.; Chang,~Y.-H.; Cheng,~Y.-J.; Li,~L.-J.
  \emph{Adv. Mater.} \textbf{2013}, \emph{25}, 3456--3461\relax
\mciteBstWouldAddEndPuncttrue
\mciteSetBstMidEndSepPunct{\mcitedefaultmidpunct}
{\mcitedefaultendpunct}{\mcitedefaultseppunct}\relax
\EndOfBibitem
\bibitem[Geim and Grigorieva(2013)Geim, and Grigorieva]{Geim:nat13}
Geim,~A.~K.; Grigorieva,~I.~V. \emph{Nature} \textbf{2013}, \emph{499},
  419--425\relax
\mciteBstWouldAddEndPuncttrue
\mciteSetBstMidEndSepPunct{\mcitedefaultmidpunct}
{\mcitedefaultendpunct}{\mcitedefaultseppunct}\relax
\EndOfBibitem
\bibitem[Lee \latin{et~al.}(2012)Lee, Zhang, Zhang, Chang, Lin, Chang, Yu,
  Wang, Chang, Li, and Lin]{Lee:advmat12}
Lee,~Y.-H.; Zhang,~X.-Q.; Zhang,~W.; Chang,~M.-T.; Lin,~C.-T.; Chang,~K.-D.;
  Yu,~Y.-C.; Wang,~J. T.-W.; Chang,~C.-S.; Li,~L.-J.; Lin,~T.-W. \emph{Adv.
  Mater.} \textbf{2012}, \emph{24}, 2320--2325\relax
\mciteBstWouldAddEndPuncttrue
\mciteSetBstMidEndSepPunct{\mcitedefaultmidpunct}
{\mcitedefaultendpunct}{\mcitedefaultseppunct}\relax
\EndOfBibitem
\bibitem[Chen \latin{et~al.}(2015)Chen, Liu, Liu, Tang, Nai, Li, Zheng, Gao,
  Zheng, Shin, Jeong, and Loh]{Chen:advmat15}
Chen,~J.; Liu,~B.; Liu,~Y.; Tang,~W.; Nai,~C.~T.; Li,~L.; Zheng,~J.; Gao,~L.;
  Zheng,~Y.; Shin,~H.~S.; Jeong,~H.~Y.; Loh,~K.~P. \emph{Adv. Mater.}
  \textbf{2015}, \emph{27}, 6722--6727\relax
\mciteBstWouldAddEndPuncttrue
\mciteSetBstMidEndSepPunct{\mcitedefaultmidpunct}
{\mcitedefaultendpunct}{\mcitedefaultseppunct}\relax
\EndOfBibitem
\bibitem[George \latin{et~al.}(2014)George, Mutlu, Ionescu, Wu, Jeong, Bay,
  Chai, Mkhoyan, Ozkan, and Ozkan]{George:afm14}
George,~A.~S.; Mutlu,~Z.; Ionescu,~R.; Wu,~R.~J.; Jeong,~J.~S.; Bay,~H.~H.;
  Chai,~Y.; Mkhoyan,~K.~A.; Ozkan,~M.; Ozkan,~C.~S. \emph{Adv. Func. Mater.}
  \textbf{2014}, \emph{24}, 7461--7466\relax
\mciteBstWouldAddEndPuncttrue
\mciteSetBstMidEndSepPunct{\mcitedefaultmidpunct}
{\mcitedefaultendpunct}{\mcitedefaultseppunct}\relax
\EndOfBibitem
\bibitem[Fang \latin{et~al.}(2012)Fang, Chuang, Chang, Takei, Takahashi, and
  Javey]{Fang:nanol12}
Fang,~H.; Chuang,~S.; Chang,~T.~C.; Takei,~K.; Takahashi,~T.; Javey,~A.
  \emph{Nano Lett.} \textbf{2012}, \emph{12}, 3788--3792\relax
\mciteBstWouldAddEndPuncttrue
\mciteSetBstMidEndSepPunct{\mcitedefaultmidpunct}
{\mcitedefaultendpunct}{\mcitedefaultseppunct}\relax
\EndOfBibitem
\bibitem[Shokouh \latin{et~al.}(2015)Shokouh, Jeon, Pezeshki, Choi, Lee, Kim,
  Park, and Im]{Shokouh:afm15}
Shokouh,~S. H.~H.; Jeon,~P.~J.; Pezeshki,~A.; Choi,~K.; Lee,~H.~S.; Kim,~J.~S.;
  Park,~E.~Y.; Im,~S. \emph{Adv. Func. Mater.} \textbf{2015},
  10.1002/adfm.201502008\relax
\mciteBstWouldAddEndPuncttrue
\mciteSetBstMidEndSepPunct{\mcitedefaultmidpunct}
{\mcitedefaultendpunct}{\mcitedefaultseppunct}\relax
\EndOfBibitem
\bibitem[Joonki \latin{et~al.}(2014)Joonki, Tae-Eon, Der-Yuh, Deyi, Joonsuk,
  Hee, Yabin, Changhyun, Chaun, Yinghui, Robert, Joonyeon, Sefaattin, and
  Junqiao]{Joonki:nanol14}
Joonki,~S.; Tae-Eon,~P.; Der-Yuh,~L.; Deyi,~F.; Joonsuk,~P.; Hee,~J.~J.;
  Yabin,~C.; Changhyun,~K.; Chaun,~J.; Yinghui,~S.; Robert,~S.; Joonyeon,~C.;
  Sefaattin,~T.; Junqiao,~W. \emph{Nano Lett.} \textbf{2014}, \emph{14},
  6976--6982\relax
\mciteBstWouldAddEndPuncttrue
\mciteSetBstMidEndSepPunct{\mcitedefaultmidpunct}
{\mcitedefaultendpunct}{\mcitedefaultseppunct}\relax
\EndOfBibitem
\bibitem[Kang \latin{et~al.}(2015)Kang, Kim, Shim, Jeon, Park, Jung, Yu, Pang,
  Lee, and Park]{Kang:afm15}
Kang,~D.-H.; Kim,~M.-S.; Shim,~J.; Jeon,~J.; Park,~H.-Y.; Jung,~W.-S.;
  Yu,~H.-Y.; Pang,~C.-H.; Lee,~S.; Park,~J.-H. \emph{Adv. Func. Mater.}
  \textbf{2015}, \emph{25}, 4219--4227\relax
\mciteBstWouldAddEndPuncttrue
\mciteSetBstMidEndSepPunct{\mcitedefaultmidpunct}
{\mcitedefaultendpunct}{\mcitedefaultseppunct}\relax
\EndOfBibitem
\bibitem[\ifmmode~\mbox{\c{C}}\else \c{C}\fi{}ak\ifmmode\imath\else\i\fi{}r
  \latin{et~al.}(2014)\ifmmode~\mbox{\c{C}}\else
  \c{C}\fi{}ak\ifmmode\imath\else\i\fi{}r, Sevik, and Peeters]{Cakir:jmcc46}
\ifmmode~\mbox{\c{C}}\else \c{C}\fi{}ak\ifmmode\imath\else\i\fi{}r,~D.;
  Sevik,~C.; Peeters,~F.~M. \emph{J. Mater. Chem. C} \textbf{2014}, \emph{2},
  9842--9849\relax
\mciteBstWouldAddEndPuncttrue
\mciteSetBstMidEndSepPunct{\mcitedefaultmidpunct}
{\mcitedefaultendpunct}{\mcitedefaultseppunct}\relax
\EndOfBibitem
\bibitem[Allain and Kis(2014)Allain, and Kis]{Allain:acs8}
Allain,~A.; Kis,~A. \emph{ACS Nano} \textbf{2014}, \emph{8}, 7180--7185\relax
\mciteBstWouldAddEndPuncttrue
\mciteSetBstMidEndSepPunct{\mcitedefaultmidpunct}
{\mcitedefaultendpunct}{\mcitedefaultseppunct}\relax
\EndOfBibitem
\bibitem[Braga \latin{et~al.}(2012)Braga, Lezama, Berger, and
  Morpurgo]{Braga:nanol12}
Braga,~D.; Lezama,~I.~G.; Berger,~H.; Morpurgo,~A.~F. \emph{Nano Lett.}
  \textbf{2012}, \emph{12}, 5218--5223\relax
\mciteBstWouldAddEndPuncttrue
\mciteSetBstMidEndSepPunct{\mcitedefaultmidpunct}
{\mcitedefaultendpunct}{\mcitedefaultseppunct}\relax
\EndOfBibitem
\bibitem[Greiner \latin{et~al.}(2013)Greiner, Chai, Helander, Tang, and
  Lu]{Greiner:afm13}
Greiner,~M.~T.; Chai,~L.; Helander,~M.~G.; Tang,~W.-M.; Lu,~Z.-H. \emph{Adv.
  Func. Mater.} \textbf{2013}, \emph{23}, 215--226\relax
\mciteBstWouldAddEndPuncttrue
\mciteSetBstMidEndSepPunct{\mcitedefaultmidpunct}
{\mcitedefaultendpunct}{\mcitedefaultseppunct}\relax
\EndOfBibitem
\bibitem[Kr{\"{o}}ger \latin{et~al.}(2009)Kr{\"{o}}ger, Hamwi, Meyer, Riedl,
  Kowalsky, and Kahn]{Kroeger:apl09}
Kr{\"{o}}ger,~M.; Hamwi,~S.; Meyer,~J.; Riedl,~T.; Kowalsky,~W.; Kahn,~A.
  \emph{Appl. Phys. Lett.} \textbf{2009}, \emph{95}, 123301\relax
\mciteBstWouldAddEndPuncttrue
\mciteSetBstMidEndSepPunct{\mcitedefaultmidpunct}
{\mcitedefaultendpunct}{\mcitedefaultseppunct}\relax
\EndOfBibitem
\bibitem[Chuang \latin{et~al.}(2014)Chuang, Battaglia, Azcatl, McDonnell, Kang,
  Yin, Tosun, Kapadia, Fang, Wallace, and Javey]{Chuang:nanol13}
Chuang,~S.; Battaglia,~C.; Azcatl,~A.; McDonnell,~S.; Kang,~J.~S.; Yin,~X.;
  Tosun,~M.; Kapadia,~R.; Fang,~H.; Wallace,~R.~M.; Javey,~A. \emph{Nano Lett.}
  \textbf{2014}, \emph{14}, 1337--1342\relax
\mciteBstWouldAddEndPuncttrue
\mciteSetBstMidEndSepPunct{\mcitedefaultmidpunct}
{\mcitedefaultendpunct}{\mcitedefaultseppunct}\relax
\EndOfBibitem
\bibitem[McDonnell \latin{et~al.}(2014)McDonnell, Addou, Buie, Wallace, and
  Hinkle]{McDonnell:acs8}
McDonnell,~S.; Addou,~R.; Buie,~C.; Wallace,~R.~M.; Hinkle,~C.~L. \emph{ACS
  Nano} \textbf{2014}, \emph{8}, 2880--2888\relax
\mciteBstWouldAddEndPuncttrue
\mciteSetBstMidEndSepPunct{\mcitedefaultmidpunct}
{\mcitedefaultendpunct}{\mcitedefaultseppunct}\relax
\EndOfBibitem
\bibitem[Chen \latin{et~al.}(2013)Chen, Santos, Zhu, Kaxiras, and
  Zhang]{Chen:nanol13}
Chen,~W.; Santos,~E. J.~G.; Zhu,~W.; Kaxiras,~E.; Zhang,~Z. \emph{Nano Lett.}
  \textbf{2013}, \emph{13}, 509--514\relax
\mciteBstWouldAddEndPuncttrue
\mciteSetBstMidEndSepPunct{\mcitedefaultmidpunct}
{\mcitedefaultendpunct}{\mcitedefaultseppunct}\relax
\EndOfBibitem
\bibitem[Farmanbar and Brocks(2015)Farmanbar, and Brocks]{Farmanbar:prb15}
Farmanbar,~M.; Brocks,~G. \emph{Phys. Rev. B} \textbf{2015}, \emph{91},
  161304\relax
\mciteBstWouldAddEndPuncttrue
\mciteSetBstMidEndSepPunct{\mcitedefaultmidpunct}
{\mcitedefaultendpunct}{\mcitedefaultseppunct}\relax
\EndOfBibitem
\bibitem[Gong \latin{et~al.}(2014)Gong, Colombo, Wallace, and
  Cho]{Gong:nanol14}
Gong,~C.; Colombo,~L.; Wallace,~R.~M.; Cho,~K. \emph{Nano Lett.} \textbf{2014},
  \emph{14}, 1714--1720\relax
\mciteBstWouldAddEndPuncttrue
\mciteSetBstMidEndSepPunct{\mcitedefaultmidpunct}
{\mcitedefaultendpunct}{\mcitedefaultseppunct}\relax
\EndOfBibitem
\bibitem[Kang \latin{et~al.}(2014)Kang, Liu, Sarkar, Jena, and
  Banerjee]{Kang:prx14}
Kang,~J.; Liu,~W.; Sarkar,~D.; Jena,~D.; Banerjee,~K. \emph{Phys. Rev. X}
  \textbf{2014}, \emph{4}, 031005\relax
\mciteBstWouldAddEndPuncttrue
\mciteSetBstMidEndSepPunct{\mcitedefaultmidpunct}
{\mcitedefaultendpunct}{\mcitedefaultseppunct}\relax
\EndOfBibitem
\bibitem[Kappera \latin{et~al.}(2014)Kappera, Voiry, Yalcin, Branch, Gupta,
  Mohite, and Chhowalla]{Chhowalla:natmat14}
Kappera,~R.; Voiry,~D.; Yalcin,~S.~E.; Branch,~B.; Gupta,~G.; Mohite,~A.~D.;
  Chhowalla,~M. \emph{Nat. Mat.} \textbf{2014}, \emph{13}, 1128--1134\relax
\mciteBstWouldAddEndPuncttrue
\mciteSetBstMidEndSepPunct{\mcitedefaultmidpunct}
{\mcitedefaultendpunct}{\mcitedefaultseppunct}\relax
\EndOfBibitem
\bibitem[Momma and Izumi(2008)Momma, and Izumi]{Momma:db5098}
Momma,~K.; Izumi,~F. \emph{J. Appl. Cryst.} \textbf{2008}, \emph{41},
  653--658\relax
\mciteBstWouldAddEndPuncttrue
\mciteSetBstMidEndSepPunct{\mcitedefaultmidpunct}
{\mcitedefaultendpunct}{\mcitedefaultseppunct}\relax
\EndOfBibitem
\bibitem[Giovannetti \latin{et~al.}(2008)Giovannetti, Khomyakov, Brocks,
  Karpan, van~den Brink, and Kelly]{Giovannetti:prl08}
Giovannetti,~G.; Khomyakov,~P.~A.; Brocks,~G.; Karpan,~V.~M.; van~den
  Brink,~J.; Kelly,~P.~J. \emph{Phys. Rev. Lett.} \textbf{2008}, \emph{101},
  026803\relax
\mciteBstWouldAddEndPuncttrue
\mciteSetBstMidEndSepPunct{\mcitedefaultmidpunct}
{\mcitedefaultendpunct}{\mcitedefaultseppunct}\relax
\EndOfBibitem
\bibitem[Khomyakov \latin{et~al.}(2009)Khomyakov, Giovannetti, Rusu, Brocks,
  van~den Brink, and Kelly]{Khomyakov:prb11}
Khomyakov,~P.~A.; Giovannetti,~G.; Rusu,~P.~C.; Brocks,~G.; van~den Brink,~J.;
  Kelly,~P.~J. \emph{Phys. Rev. B} \textbf{2009}, \emph{79}, 195425\relax
\mciteBstWouldAddEndPuncttrue
\mciteSetBstMidEndSepPunct{\mcitedefaultmidpunct}
{\mcitedefaultendpunct}{\mcitedefaultseppunct}\relax
\EndOfBibitem
\bibitem[Bokdam \latin{et~al.}(2014)Bokdam, Brocks, Katsnelson, and
  Kelly]{Bokdam:prb14a}
Bokdam,~M.; Brocks,~G.; Katsnelson,~M.~I.; Kelly,~P.~J. \emph{Phys. Rev. B}
  \textbf{2014}, \emph{90}, 085415\relax
\mciteBstWouldAddEndPuncttrue
\mciteSetBstMidEndSepPunct{\mcitedefaultmidpunct}
{\mcitedefaultendpunct}{\mcitedefaultseppunct}\relax
\EndOfBibitem
\bibitem[Chuang \latin{et~al.}(2014)Chuang, Tan, Ghimire, Perera, Chamlagain,
  Cheng, Yan, Mandrus, Tom{\`{a}}nek, and Zhou]{Chuang:nanol14}
Chuang,~H.-J.; Tan,~X.; Ghimire,~N.~J.; Perera,~M.~M.; Chamlagain,~B.;
  Cheng,~M. M.-C.; Yan,~J.; Mandrus,~D.; Tom{\`{a}}nek,~D.; Zhou,~Z. \emph{Nano
  Lett.} \textbf{2014}, \emph{14}, 3594--3601\relax
\mciteBstWouldAddEndPuncttrue
\mciteSetBstMidEndSepPunct{\mcitedefaultmidpunct}
{\mcitedefaultendpunct}{\mcitedefaultseppunct}\relax
\EndOfBibitem
\bibitem[Leong \latin{et~al.}(2015)Leong, Luo, Li, Khoo, Quek, and
  Thong]{Leong:acs15}
Leong,~W.~S.; Luo,~X.; Li,~Y.; Khoo,~K.~H.; Quek,~S.~Y.; Thong,~J. T.~L.
  \emph{ACS Nano} \textbf{2015}, \emph{9}, 869--877\relax
\mciteBstWouldAddEndPuncttrue
\mciteSetBstMidEndSepPunct{\mcitedefaultmidpunct}
{\mcitedefaultendpunct}{\mcitedefaultseppunct}\relax
\EndOfBibitem
\bibitem[Mattheiss(1973)]{Mattheis:prb73}
Mattheiss,~L.~F. \emph{Phys. Rev. B} \textbf{1973}, \emph{8}, 3719--3740\relax
\mciteBstWouldAddEndPuncttrue
\mciteSetBstMidEndSepPunct{\mcitedefaultmidpunct}
{\mcitedefaultendpunct}{\mcitedefaultseppunct}\relax
\EndOfBibitem
\bibitem[Lince \latin{et~al.}(1987)Lince, Carr\'e, and
  Fleischauer]{Lince:prb87}
Lince,~J.~R.; Carr\'e,~D.~J.; Fleischauer,~P.~D. \emph{Phys. Rev. B}
  \textbf{1987}, \emph{36}, 1647--1656\relax
\mciteBstWouldAddEndPuncttrue
\mciteSetBstMidEndSepPunct{\mcitedefaultmidpunct}
{\mcitedefaultendpunct}{\mcitedefaultseppunct}\relax
\EndOfBibitem
\bibitem[Yazyev and Pasquarello(2009)Yazyev, and Pasquarello]{Yazyev:prb09}
Yazyev,~O.; Pasquarello,~A. \emph{Phys. Rev. B} \textbf{2009}, \emph{80},
  035408\relax
\mciteBstWouldAddEndPuncttrue
\mciteSetBstMidEndSepPunct{\mcitedefaultmidpunct}
{\mcitedefaultendpunct}{\mcitedefaultseppunct}\relax
\EndOfBibitem
\bibitem[Bokdam \latin{et~al.}(2014)Bokdam, Brocks, and Kelly]{Bokdam:prb14b}
Bokdam,~M.; Brocks,~G.; Kelly,~P.~J. \emph{Phys. Rev. B} \textbf{2014},
  \emph{90}, 201411\relax
\mciteBstWouldAddEndPuncttrue
\mciteSetBstMidEndSepPunct{\mcitedefaultmidpunct}
{\mcitedefaultendpunct}{\mcitedefaultseppunct}\relax
\EndOfBibitem
\bibitem[Li \latin{et~al.}(2014)Li, Komatsu, Nakaharai, Lin, Yamamoto, Duan,
  and Tsukagoshi]{Song:ACS8}
Li,~S.-L.; Komatsu,~K.; Nakaharai,~S.; Lin,~Y.-F.; Yamamoto,~M.; Duan,~X.;
  Tsukagoshi,~K. \emph{ACS Nano} \textbf{2014}, \emph{8}, 12836--12842\relax
\mciteBstWouldAddEndPuncttrue
\mciteSetBstMidEndSepPunct{\mcitedefaultmidpunct}
{\mcitedefaultendpunct}{\mcitedefaultseppunct}\relax
\EndOfBibitem
\bibitem[Britnell \latin{et~al.}(2012)Britnell, Gorbachev, Jalil, Belle,
  Schedin, Katsnelson, Eaves, Morozov, Mayorov, Peres, Neto, Leist, Geim,
  Ponomarenko, and Novoselov]{Britnell:nanol12}
Britnell,~L.; Gorbachev,~R.~V.; Jalil,~R.; Belle,~B.~D.; Schedin,~F.;
  Katsnelson,~M.~I.; Eaves,~L.; Morozov,~S.~V.; Mayorov,~A.~S.; Peres,~N.
  M.~R.; Neto,~A. H.~C.; Leist,~J.; Geim,~A.~K.; Ponomarenko,~L.~A.;
  Novoselov,~K.~S. \emph{Nano Lett.} \textbf{2012}, \emph{12}, 1707--1710\relax
\mciteBstWouldAddEndPuncttrue
\mciteSetBstMidEndSepPunct{\mcitedefaultmidpunct}
{\mcitedefaultendpunct}{\mcitedefaultseppunct}\relax
\EndOfBibitem
\bibitem[Kresse and Hafner(1993)Kresse, and Hafner]{Kresse:prb93}
Kresse,~G.; Hafner,~J. \emph{Phys. Rev. B} \textbf{1993}, \emph{47},
  558--561\relax
\mciteBstWouldAddEndPuncttrue
\mciteSetBstMidEndSepPunct{\mcitedefaultmidpunct}
{\mcitedefaultendpunct}{\mcitedefaultseppunct}\relax
\EndOfBibitem
\bibitem[Kresse and Furthm{\"{u}}ller(1996)Kresse, and
  Furthm{\"{u}}ller]{Kresse:prb96}
Kresse,~G.; Furthm{\"{u}}ller,~J. \emph{Phys. Rev. B} \textbf{1996}, \emph{54},
  11169--11186\relax
\mciteBstWouldAddEndPuncttrue
\mciteSetBstMidEndSepPunct{\mcitedefaultmidpunct}
{\mcitedefaultendpunct}{\mcitedefaultseppunct}\relax
\EndOfBibitem
\bibitem[Bl{\"{o}}chl(1994)]{Blochl:prb94b}
Bl{\"{o}}chl,~P.~E. \emph{Phys. Rev. B} \textbf{1994}, \emph{50},
  17953--17979\relax
\mciteBstWouldAddEndPuncttrue
\mciteSetBstMidEndSepPunct{\mcitedefaultmidpunct}
{\mcitedefaultendpunct}{\mcitedefaultseppunct}\relax
\EndOfBibitem
\bibitem[Kresse and Joubert(1999)Kresse, and Joubert]{Kresse:prb99}
Kresse,~G.; Joubert,~D. \emph{Phys. Rev. B} \textbf{1999}, \emph{59},
  1758--1775\relax
\mciteBstWouldAddEndPuncttrue
\mciteSetBstMidEndSepPunct{\mcitedefaultmidpunct}
{\mcitedefaultendpunct}{\mcitedefaultseppunct}\relax
\EndOfBibitem
\bibitem[Farmanbar and Brocks(2015)Farmanbar, and Brocks]{Farmanbar:arxiv}
Farmanbar,~M.; Brocks,~G. \emph{arXiv:1510.04337v2} \textbf{2015}, \relax
\mciteBstWouldAddEndPunctfalse
\mciteSetBstMidEndSepPunct{\mcitedefaultmidpunct}
{}{\mcitedefaultseppunct}\relax
\EndOfBibitem
\bibitem[B\"oker \latin{et~al.}(2001)B\"oker, Severin, M\"uller, Janowitz,
  Manzke, Vo\ss{}, Kr\"uger, Mazur, and Pollmann]{Boker:prb01}
B\"oker,~T.; Severin,~R.; M\"uller,~A.; Janowitz,~C.; Manzke,~R.; Vo\ss{},~D.;
  Kr\"uger,~P.; Mazur,~A.; Pollmann,~J. \emph{Phys. Rev. B} \textbf{2001},
  \emph{64}, 235305\relax
\mciteBstWouldAddEndPuncttrue
\mciteSetBstMidEndSepPunct{\mcitedefaultmidpunct}
{\mcitedefaultendpunct}{\mcitedefaultseppunct}\relax
\EndOfBibitem
\bibitem[Schutte \latin{et~al.}(1987)Schutte, Boer, and
  Jellinek]{Schutte:jssc87}
Schutte,~W.; Boer,~J.~D.; Jellinek,~F. \emph{J. Solid State Chem.}
  \textbf{1987}, \emph{70}, 207 -- 209\relax
\mciteBstWouldAddEndPuncttrue
\mciteSetBstMidEndSepPunct{\mcitedefaultmidpunct}
{\mcitedefaultendpunct}{\mcitedefaultseppunct}\relax
\EndOfBibitem
\bibitem[Ohno(1998)]{Ohno:prb98}
Ohno,~Y. \emph{Phys. Rev. B} \textbf{1998}, \emph{58}, 8042--8049\relax
\mciteBstWouldAddEndPuncttrue
\mciteSetBstMidEndSepPunct{\mcitedefaultmidpunct}
{\mcitedefaultendpunct}{\mcitedefaultseppunct}\relax
\EndOfBibitem
\bibitem[Chiu \latin{et~al.}(2015)Chiu, Zhang, Shiu, Chuu, Chen, Chang, Chen,
  Chou, Shih, and Li]{Chiu:natcom15}
Chiu,~M.-H.; Zhang,~C.; Shiu,~H.-W.; Chuu,~C.-P.; Chen,~C.-H.; Chang,~C.-Y.~S.;
  Chen,~C.-H.; Chou,~M.-Y.; Shih,~C.-K.; Li,~L.-J. \emph{Nat. Commun.}
  \textbf{2015}, \emph{6}, 7666\relax
\mciteBstWouldAddEndPuncttrue
\mciteSetBstMidEndSepPunct{\mcitedefaultmidpunct}
{\mcitedefaultendpunct}{\mcitedefaultseppunct}\relax
\EndOfBibitem
\bibitem[Mak \latin{et~al.}(2010)Mak, Lee, Hone, Shan, and Heinz]{Mak:prl10}
Mak,~K.~F.; Lee,~C.; Hone,~J.; Shan,~J.; Heinz,~T.~F. \emph{Phys. Rev. Lett.}
  \textbf{2010}, \emph{105}, 136805\relax
\mciteBstWouldAddEndPuncttrue
\mciteSetBstMidEndSepPunct{\mcitedefaultmidpunct}
{\mcitedefaultendpunct}{\mcitedefaultseppunct}\relax
\EndOfBibitem
\bibitem[Tongay \latin{et~al.}(2012)Tongay, Zhou, Ataca, Lo, Matthews, Li,
  Grossman, and Wu]{Tongay:nanolett12}
Tongay,~S.; Zhou,~J.; Ataca,~C.; Lo,~K.; Matthews,~T.~S.; Li,~J.;
  Grossman,~J.~C.; Wu,~J. \emph{Nano Lett.} \textbf{2012}, \emph{12},
  5576--5580\relax
\mciteBstWouldAddEndPuncttrue
\mciteSetBstMidEndSepPunct{\mcitedefaultmidpunct}
{\mcitedefaultendpunct}{\mcitedefaultseppunct}\relax
\EndOfBibitem
\bibitem[Ugeda \latin{et~al.}(2014)Ugeda, Bradley, Shi, da~Jornada, Zhang, Qiu,
  Ruan, Mo, Hussain, Shen, Wang, Louie, and Crommie]{Ugeda:natmat14}
Ugeda,~M.~M.; Bradley,~A.~J.; Shi,~S.-F.; da~Jornada,~F.~H.; Zhang,~Y.;
  Qiu,~D.~Y.; Ruan,~W.; Mo,~S.-K.; Hussain,~Z.; Shen,~Z.-X.; Wang,~F.;
  Louie,~S.~G.; Crommie,~M.~F. \emph{Nat. mater.} \textbf{2014}, \emph{13},
  1091\relax
\mciteBstWouldAddEndPuncttrue
\mciteSetBstMidEndSepPunct{\mcitedefaultmidpunct}
{\mcitedefaultendpunct}{\mcitedefaultseppunct}\relax
\EndOfBibitem
\bibitem[Ruppert \latin{et~al.}(2014)Ruppert, Aslan, and
  Heinz]{Ruppert:nanolett12}
Ruppert,~C.; Aslan,~O.~B.; Heinz,~T.~F. \emph{Nano Lett.} \textbf{2014},
  \emph{14}, 6231--6236\relax
\mciteBstWouldAddEndPuncttrue
\mciteSetBstMidEndSepPunct{\mcitedefaultmidpunct}
{\mcitedefaultendpunct}{\mcitedefaultseppunct}\relax
\EndOfBibitem
\bibitem[Guti{\'{e}}rrez \latin{et~al.}(2013)Guti{\'{e}}rrez,
  Perea-L{\'{o}}pez, El{\'{i}}­as, Berkdemir, Wang, Lv,
  L{\'{o}}pez-Ur{\'{i}}­as, Crespi, Terrones, and
  Terrones]{Gutierrez:nanolett13}
Guti{\'{e}}rrez,~H.~R.; Perea-L{\'{o}}pez,~N.; El{\'{i}}­as,~A.~L.;
  Berkdemir,~A.; Wang,~B.; Lv,~R.; L{\'{o}}pez-Ur{\'{i}}­as,~F.; Crespi,~V.~H.;
  Terrones,~H.; Terrones,~M. \emph{Nano Lett.} \textbf{2013}, \emph{13},
  3447--3454\relax
\mciteBstWouldAddEndPuncttrue
\mciteSetBstMidEndSepPunct{\mcitedefaultmidpunct}
{\mcitedefaultendpunct}{\mcitedefaultseppunct}\relax
\EndOfBibitem
\bibitem[Dion \latin{et~al.}(2004)Dion, Rydberg, Schr{\"{o}}der, Langreth, and
  Lundqvist]{Dion:prl04}
Dion,~M.; Rydberg,~H.; Schr{\"{o}}der,~E.; Langreth,~D.~C.; Lundqvist,~B.~I.
  \emph{Phys. Rev. Lett.} \textbf{2004}, \emph{92}, 246401\relax
\mciteBstWouldAddEndPuncttrue
\mciteSetBstMidEndSepPunct{\mcitedefaultmidpunct}
{\mcitedefaultendpunct}{\mcitedefaultseppunct}\relax
\EndOfBibitem
\bibitem[Klime{\v{s}} \latin{et~al.}(2011)Klime{\v{s}}, Bowler, and
  Michaelides]{Klimes:prb11}
Klime{\v{s}},~J.; Bowler,~D.~R.; Michaelides,~A. \emph{Phys. Rev. B}
  \textbf{2011}, \emph{83}, 195131\relax
\mciteBstWouldAddEndPuncttrue
\mciteSetBstMidEndSepPunct{\mcitedefaultmidpunct}
{\mcitedefaultendpunct}{\mcitedefaultseppunct}\relax
\EndOfBibitem
\bibitem[Thonhauser \latin{et~al.}(2007)Thonhauser, Cooper, Li, Puzder,
  Hyldgaard, and Langreth]{Thonhauser:prb76}
Thonhauser,~T.; Cooper,~V.~R.; Li,~S.; Puzder,~A.; Hyldgaard,~P.;
  Langreth,~D.~C. \emph{Phys. Rev. B} \textbf{2007}, \emph{76}, 125112\relax
\mciteBstWouldAddEndPuncttrue
\mciteSetBstMidEndSepPunct{\mcitedefaultmidpunct}
{\mcitedefaultendpunct}{\mcitedefaultseppunct}\relax
\EndOfBibitem
\bibitem[Dawson and Bullett(1987)Dawson, and Bullett]{Dawson:jpc87}
Dawson,~W.~G.; Bullett,~D.~W. \emph{J. Phys. C: Solid State Phys.}
  \textbf{1987}, \emph{20}, 6159\relax
\mciteBstWouldAddEndPuncttrue
\mciteSetBstMidEndSepPunct{\mcitedefaultmidpunct}
{\mcitedefaultendpunct}{\mcitedefaultseppunct}\relax
\EndOfBibitem
\bibitem[van~der Horst \latin{et~al.}(1999)van~der Horst, Bobbert, Michels,
  Brocks, and Kelly]{vanderhorst:prl99}
van~der Horst,~J.-W.; Bobbert,~P.~A.; Michels,~M. A.~J.; Brocks,~G.;
  Kelly,~P.~J. \emph{Phys. Rev. Lett.} \textbf{1999}, \emph{83},
  4413--4416\relax
\mciteBstWouldAddEndPuncttrue
\mciteSetBstMidEndSepPunct{\mcitedefaultmidpunct}
{\mcitedefaultendpunct}{\mcitedefaultseppunct}\relax
\EndOfBibitem
\bibitem[Zhang \latin{et~al.}(2014)Zhang, Chang, Zhou, Cui, Yan, Liu, Schmitt,
  Lee, Moore, Chen, Lin, Jeng, Mo, Hussain, Bansil, and Shen]{Zhang:natnano14}
Zhang,~Y. \latin{et~al.}  \emph{Nat. Nanotechnol.} \textbf{2014}, \emph{9},
  111\relax
\mciteBstWouldAddEndPuncttrue
\mciteSetBstMidEndSepPunct{\mcitedefaultmidpunct}
{\mcitedefaultendpunct}{\mcitedefaultseppunct}\relax
\EndOfBibitem
\bibitem[Kang \latin{et~al.}(2013)Kang, Tongay, Zhou, Li, and Wu]{Kang:apl13}
Kang,~J.; Tongay,~S.; Zhou,~J.; Li,~J.; Wu,~J. \emph{Appl. Phys. Lett.}
  \textbf{2013}, \emph{102}, 012111\relax
\mciteBstWouldAddEndPuncttrue
\mciteSetBstMidEndSepPunct{\mcitedefaultmidpunct}
{\mcitedefaultendpunct}{\mcitedefaultseppunct}\relax
\EndOfBibitem
\end{mcitethebibliography}
\end{document}